# Quasi 1D MoS$_2$ Nanoribbons with Enhanced Edge Nonlinear Response and Photoresponsivity

Ganesh Ghimire,[†] Rajesh Kumar Ulaganathan,[†] Agnès Tempez,[‡] Oleksii Ilchenko,[¶] Raymond R. Unocic,[§] Julian Heske,[∥] Denys I. Miakota,[†] Cheng Xiang,[∥] Marc Chaigneau,[‡] Tim Booth,[∥] Peter Bøggild,[∥] Kristian S. Thygesen,[∥] David B. Geohegan,[§] and Stela Canulescu[*,†]

[†]*Department of Electrical and Photonics Engineering, Technical University of Denmark, 4000 Roskilde, Denmark*
[‡]*Horiba France SAS, Palaiseau, France*
[¶]*Department of Health Technology Nanoprobes, Technical University of Denmark, 2800 Kgs. Lyngby, Denmark*
[§]*Center for Nanophase Materials Sciences and Materials Science and Technology Division, Oak Ridge National Laboratory, Oak Ridge, 37830 Tennessee, United States*
[∥]*Department of Physics, Technical University of Denmark, 2800 Kgs. Lyngby, Denmark*

E-mail: stec@fotonik.dtu.dk

## Abstract

MoS$_2$ nanoribbons have attracted increased interest due to their properties which can be tailored by tuning their dimensions. Herein, we demonstrate the growth of quasi-one-dimensional (1D) MoS$_2$ nanoribbons and 3D triangular crystals formed by the reaction between ultra-thin films of MoO$_3$ grown by Pulsed Laser Deposition and NaF in a sulfur-rich environment. The nanoribbons can reach up to 10 $\mu$m in length, and feature single-layer edges, thereby forming a single-multilayer homojunction. The single-layer edges of the nanoribbons show a pronounced second harmonic generation due to the symmetry breaking, in contrast to the centrosymmetric multilayer structure, which is unsusceptible to the second-order nonlinear process. A pronounced splitting of the Raman spectra was observed in quasi-1D $MoS_2$ nanoribbons arising from distinct contributions from the monolayer edges and multilayer core. Nanoscale imaging reveals a blue-shifted exciton emission of the monolayer edge compared to the individual triangular $MoS_2$ monolayers due to built-in local strain and disorder. We further report on an ultrasensitive photodetector made of a single quasi-1D $MoS_2$ nanoribbon with a responsivity of 8.72×$10^2$ A/W, which is among the highest reported up-to-date for single-nanoribbon photodetectors. Our findings can inspire the design of $MoS_2$ semiconductors with tunable geometries for efficient optoelectronic devices.

## Introduction

One-dimensional (1D) semiconductor nanostructures, such as nanorods, nanowires, nanobelts and nanoribbons, have been predicted to exhibit strikingly distinct functionalities from their 2D counterpart. Due to their high surface-to-volume ratios, the 1D semiconductor nanostructures have emerged as key components for various applications, including nanolasers,[1] light-emitting diodes[2] and photodetectors.[3] In particular, photodetectors based on 1D semiconductor nanoribbons, including GaSe,[4] ZrGeTe,[3] with various bandgaps in the visible light spectral region and broad spectral responses have been reported. Among two-dimensional (2D) materials, transition metal dichalcogenides (TMDs), such as $MoS_2$, have sparked a great interest due to the possibility of tailoring their electrical, optical and magnetic properties via substitutional,[5] electrostatic doping,[6] and charge transfer.[7] The excellent photon absorption and emission of single-layer TMDs, fast photocurrent switching and high photoresponsivity have sparked a great interest to realise photodetectors,[8] light-emitting diodes,[9] and solar cells.[10] Recently, new exotic properties have been realised in TMDs with reduced dimensionality. WS$_2$ nanorods with mere broken symmetry have shown an enhanced shift current and bulk photovoltaic effect.[11] MoS$_2$ nanoribbons with S-terminated zigzag edges have been theoretically predicted to have high thermodynamic stability and edge-dependent characteristics, with armchair and zigzag edges exhibiting semiconducting and metallic character, respectively.[12,13] Moreover, the optical, electrical, magnetic and catalytic properties of TMD nanoribbons can be modified in the presence of edge defects, [14] by doping,[15] strain,[16] and strain-induced polarization.[17] Hence, exploring facile methods for the synthetic design of TMD nanoribbons with tailored dimensionality can open new avenues for realising nanodevices with novel functionalities.

The 1D MoS$_2$ nanoribbon geometry has been previously achieved using patterned templates or the focused ion beam milling method.[18,19] Alternatively, a bottom-up synthesis approach can be adopted based on vapour–liquid–solid (VLS) approach, which has been well established for 1D growth of various materials, including graphene[20] and GaSe nanoribbons.[21] Li et al. have demonstrated the growth of MoS$_2$ nanoribbons using Ni particles as promoters. [22] Wu et al. reported on the growth of MoS$_2$ nanoribbons on sapphire without catalysts.[23] In the VLS growth, the super-saturated liquid droplets are formed by a sodium chloride (NaCl) reaction with molybdenum oxide (MoO$_3$).[24] Furthermore, the VLS method can be used for one-dimensional (1D) growth because 1D nanostructures are favourable during the growth steps from super-saturated catalytic droplets. The substrate can also influence growth to a large extent, which is one inherent restriction of this approach. Similarly, Pho et al. have demonstrated a 1D molybdenum diselenide (MoSe$_2$) nanoribbons by molecular beam epitaxy (MBE) on highly aligned polycrystalline graphite.[25] Cheng et al. reported the growth of MoSe$_2$ nanoribbons using a patterned gold (Au-100) substrate.[26] Despite advances in the synthetic methods, scarce reports on photodetectors based on TMDs nanoribbon arrays have only shown modest photodetectivity MoS$_2$.[4,18]

We herein demonstrate an ultra-sensitive photodetector based on a single quasi-1D MoS$_2$

nanoribbon fabricated by a VLS process. Our approach involves the reaction between uniform oxide precursors grown by pulsed laser deposition (PLD) and NaF in an S-rich environment. The synthesis method yields high-quality $MoS_2$ nanostructures consisting of quasi-1D nanoribbons and 2D triangular crystals and aligned 3D crystals with either 3R or 2H stacking orientation. Controlling the growth parameters, namely temperature and oxide precursor thickness, allows quasi-1D $MoS_2$ nanoribbons to reach several micrometres in length and tens of nanometers in height. Moreover, the quasi-1D $MoS_2$ nanoribbons exhibit monolayer edges, as revealed by scanning electron microscopy (SEM), atomic force microscopy (AFM), and second harmonic generation (SHG). Tip-enhanced photoluminescence (TEPL) spectroscopy reveals distinct PL features originating from monolayer nanoribbons and 2D crystals, with nanoribbons exhibiting blue-shifted PL emission compared to 2D $MoS_2$ crystals, owing to the built-in local strain in the nanoribbon. In addition, we report on the first photodetector based on a single quasi-1D $MoS_2$ nanoribbon on rigid substrates ($SiO_2$/Si). The response generated under illumination is orders of magnitude larger than other nanoribbon materials, owing to the high crystallinity of the $MoS_2$ nanostructures. Our findings underline the great promise of TMD-based nanoribbons for future applications in next-generation electronics and optoelectronic devices.

## Results and discussion

**Synthesis of quasi-1D $MoS_2$ nanoribbons.** The $MoS_2$ nanoribbons were grown in a two-step process in which ultra-thin oxide films were sulfurised in the presence of alkali halide promoters. The role of alkali metal halides in promoting the unidirectional synthesis of $MoS_2$ nanoribbons from $MoO_3$ solid precursor has been discussed in our previous work.[27] Notably, in the absence of the alkali metal halide layer promoter (NaF), the oxide-to-sulfide conversion leads to quasi-continuous $MoS_2$ layers of quality like the conventional CVD process (see Figure S1). Figure S2 describes the details of the synthesis process. Briefly, ultra-thin films of molybdenum oxide $MoO_x$ (2<x<3) grown by pulsed laser deposition (PLD) on c-plane sapphire serve as precursors. PLD is a versatile tool that allows a wide tunability of the precursor uniformity, thickness and stoichiometry, the latter by varying the oxygen content[28,29]. In the second step, the precursor oxide films coated with alkali metal halide (NaF) are sulfurised in an S-rich atmosphere. In contrast to the conventional CVD, our approach allows decoupling evaporation of the growth processes. At a low temperature ($600^0C$), the uniform $MoO_x$ precursor thin-film reacts with NaF to form liquid Na–Mo–O phases, such as $Na_2MoO_4$ or $Na_2Mo_2O_7$ and volatile molybdenum oxyhalides phases, such as $NaMo_2F_2$. The resulting products have much lower melting temperatures than starting solid-state precursors of $MoO_2$, $MoO_3$ or NaF (see Supplementary section 4 for details). It is important to note that both liquid and gaseous precursors contribute to the growth of $MoS_2$, from the liquid phase via the vapour-liquid-solid (VLS) process[27,30] and gas-phase via the vapour-solid (VS) mechanism process,[31,32] respectively. When sulfur is introduced at a high temperature ($800^0C$), it dissolves into a liquid droplet forming $MoS_2$ nucleation sites via the VVS process. As $MoS_2$ continues to precipitate, the liquid droplet will expand on the sapphire surface leading to unidirectional nanoribbon growth.[27,30] During the lateral growth of the nanoribbons on sapphire, the sulfur atoms (predominantly thermally cracked $S_2$ molecules[33]) will continue to dissolve into the Na-Mo-O liquid phase, leading to a 2D growth of $MoS_2$ from the edges of the nanoribbon. Indeed, a series of start/stop experiments, in which the synthesis was intentionally interrupted at the temperature range from 600 to $800^0C$ reveal that the 1D nanostructures are already formed at $600^0C$. Still, the monolayer edges are barely visible at the low temperature and very pronounced at $800^0C$ (Figure S6). The crystal symmetry of the multilayer nanoribbon presumably defines the growth direction of the 2D-$MoS_2$ from the nanoribbon edges. Occasionally, regions with lower precursor concentration lead to the growth of triangular-shaped $MoS_2$ crystals. Lastly, the volatile gaseous precursors of $NaMo_2F_2$ generated in the reaction will supply an increased concentration of the Mo precursor during growth, facilitating the growth of multilayer $MoS_2$ in both 1D- and 3D-structures.[32] As discussed later, the quasi-1D nanoribbons feature monolayer edges and, hence, single layer (1L)-multilayer (ML) homojunction characteristics due to the abrupt discontinuity in thickness and thus band gap.

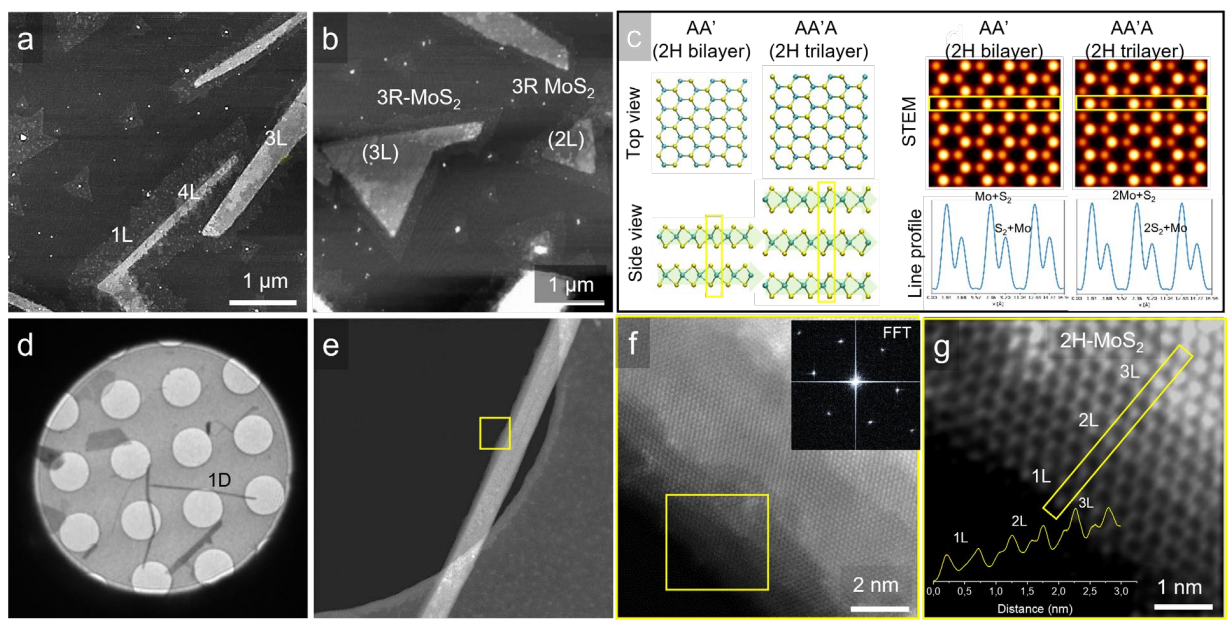

Figure 1: **Morphology and atomic resolution images of the quasi-1D MoS₂ nanoribbon.** (a) AFM image of the as-grown multilayer quasi-1D nanoribbons with lengths larger than 6 *μ*m and single-layer nanoribbon edges. (b) AFM image of the bilayer (2L) and trilayer (3L) MoS₂ triangular crystals with 3R stacking orientation. (c) Relaxed atomic models of the 2H bilayer (AA´ stacking) and trilayer (AA´A stacking) MoS₂, with top and side views, respectively. Simulated STEM images of the bilayer and 2H trilayer 2H-MoS₂ based on the theoretical relaxed structures with the same beam energy, convergence angle, and collection angles as in data acquisition. Line scans across the lattice points indicated by the overlaid rectangles. (d) TEM image of the 1D structures transferred on the TEM grid. (e-g) High-resolution ADF-STEM images of the nanoribbon edge. The inset in (f) shows the Fast Fourier Transform (FFT) spectrum of the STEM image. Experimental line profiles across the rectangle overlaid in (g) indicate a 2H stacking of the 2L and 3L quasi-1D MoS₂.

Figure 1a shows the AFM images of the quasi-1D $MoS_2$ nanoribbons on sapphire. The nanostructures are primarily multilayer nanoribbons with a thickness ranging between 2 and 3 nm, corresponding to three (3L) and four layers (4L) $MoS_2$, respectively. They have lengths of up to 10 µm and widths of up to 500 nm, yielding length-to-width ratios of 20 (see Supporting Information, Figures S3, S4). Herein, we will refer to these nanostructures as quasi-1D nanoribbons to reflect their reduced dimensionality in one direction compared to the triangle-shaped crystals. Remarkably, the quasi-1D nanoribbons exhibit monolayer edges that extend over several hundred nanometers, as can be unambiguously distinguished in the AFM line profiles shown in Figure S5. The monolayer nanoribbon is a single-crystal film that emerged upon the coalescence of multiple adjacent $MoS_2$ domains, as indicated by the overlaid triangles in Figure 1b. Depending on the precise manner two $MoS_2$ domains merge, various types of grain boundaries (GBs) consisting of 4|4, 5|7, 6|8, and 4|6 dislocation core structures can form, which have been predicted by previous theoretical studies to hamper the properties of devices.[34] Here, the obdetailsion of zigzag edges with $60^0$ angle implies that single-crystal monolayer nanoribbons are formed from highly-aligned $MoS_2$ domains with $0^0$ orientation angle. The multilayer nanoribbon core presumably drives the lateral growth of self-aligned nanoribbons.

Besides the quasi-1D structures, $MoS_2$ triangles were frequently observed on the c-plane sapphire substrate.[35] Figure 1c shows the AFM images of single layers, as well as bi- and trilayer triangles with 2H stacking orientation, namely AA′ and AA′A, respectively. Notably, various annotations are being used in the literature for the 2H (space group $P6_3/mmc$), and 3R (space group R3m) stacking prototypes of $MoS_2$.[36,37] Here, we adopt the spectroscopic notations defined in theoretical studies[38,39] for which AA′ reflects a 2H-$MoS_2$ bilayer with $180^0$ twist angle between layers, restoring the inversion symmetry. In contrast to single-layer 2H-MoS$_2$, the 2H stacking orientation restores the inversion symmetry in bulk.[37] Hence, the symbol prime in the aforementioned notation designates a mirror symmetry. This is in strong contrast to the 3R-stacked MoS$_2$, in which the layers share the same crystallographic orientation and shifts relative to the bottom layer by $\sqrt{3}a/3$ along the ZZ direction, leading to an ABC stacking order.

Next, the stacking order of the multilayer MoS$_2$ nanoribbons was explored using atomic-resolution annular dark field (ADF) imaging on an aberration-corrected scanning transmission electron microscope (STEM). For this purpose, the as-grown specimens were transferred onto a TEM grid (see Methods and Figure 1e). The image intensity is proportional to the number of layers, providing an easy and accurate way to measure the thickness of MoS$_2$ layers for thin samples, as demonstrated previously for PLD-grown films MoS$_2$.[34] Figure 1(e-h) shows the ADF-STEM images of the multilayer nanoribbons. The layered structure is evidenced in the atomic resolution image shown in Figure 1g. A magnified view over the edge of the MoS$_2$ nanoribbon reveals an image intensity contrast due to a change in the number of layers. The atom-by-atom analysis of the image intensity enables us to identify the intensity profiles from 1L, 2L, and 3L MoS$_2$. The fast Fourier transform (FFT) of the multilayer nanoribbon shown in the inset of Figure 1g illustrates one set of 6-fold diffraction points associated with the hexagonal crystal structure, which indicates a high-quality epitaxial multilayer MoS$_2$ nanoribbon with either 2H or 3H stacking orientation. To correctly assess the stacking order of the multilayer systems, we performed first-principles calculations based on density functional theory (DFT). Nonetheless, in this case, the 3R stacking registry can be ruled out due to the absence of the S$_2$ atoms in the honeycomb lattice (see simulated STEM images in Figure S7). The atomic structures of the 2H AA$'$ bilayer and AA$'$A trilayer are shown in Figure 1d with the relaxed structural models extracted from the DFT calculations. The simulated STEM images show distinct intensities at two sublattices, that is, Mo+S$_2$, S$_2$+Mo for bilayer MoS$_2$ (AA$'$ stacking) and 2Mo+S$_2$, 2S$_2$+Mo for trilayer MoS$_2$ (AA$'$A stacking). The line scans across two inequivalent lattice points in the simulated

STEM images reveal an intensity ratio difference of around 8% between the 2H-bilayer and trilayer $MoS_2$. The experimental line profiles shown in Figure 1h agree with the theoretical findings. However, we note that the experimental images are less sharp than the simulated ones due to electron probe tailing and drifting effects during STEM imaging. Lastly, while the 2H stacking prevails over the 3R stacking, we cannot exclude the presence of both low-energy stacking sequences.[39]

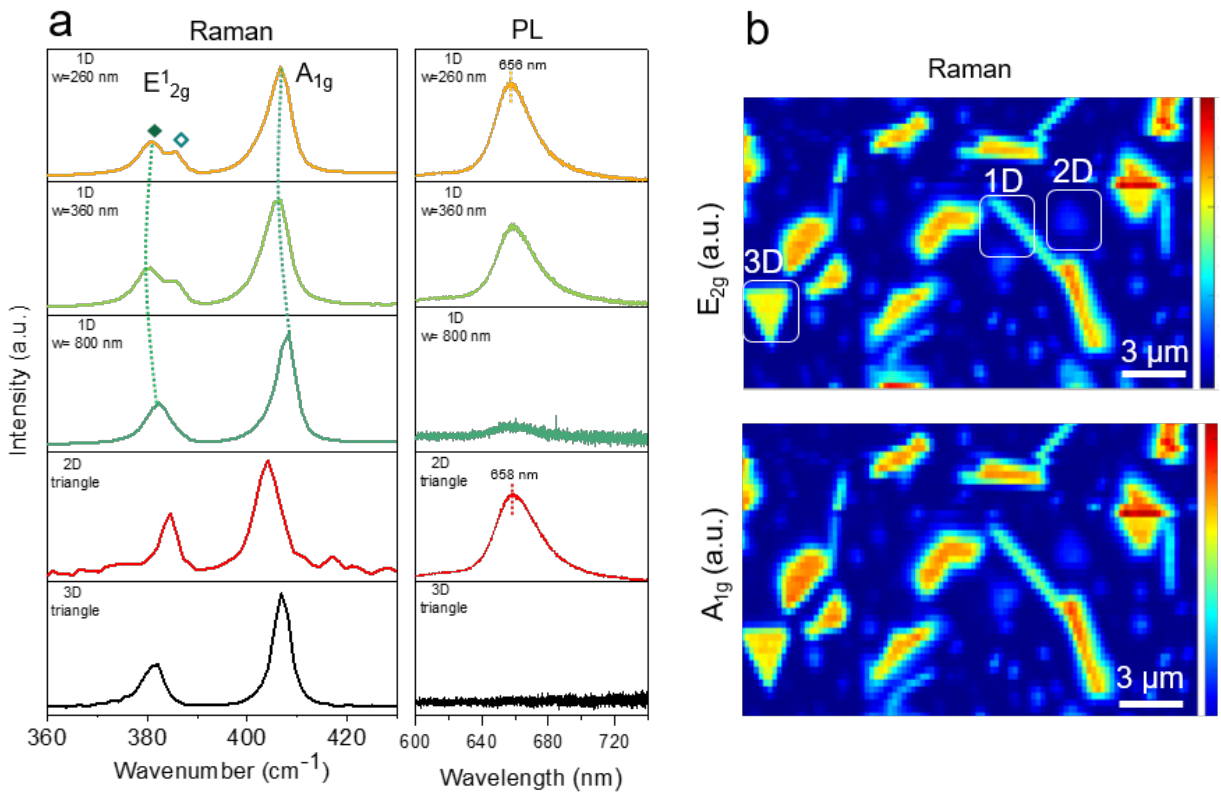

Figure 2: **Optical properties of the $MoS_2$ nanostructures with distinct dimensions**. (a) Raman and PL spectra of triangle-shaped monolayer (1L), quasi-1D nanoribbon (1D) featuring nanoribbon core widths of 260, 360, and 800 nm, and multilayer triangular (3D) $MoS_2$ crystals. (b) Raman maps showing the intensity of $E^1_2$ peak (top image) and $A_{1g}$ peak (bottom image). The full and empty diamond symbols denote the Raman peaks of the core and monolayer nanoribbon, respectively. The parameter $w$ indicates the width of the multilayer nanoribbon core, excluding the contribution from the monolayer nanoribbon edge.

**Optical properties of the quasi-1D MoS₂ nanoribbon.** We have then explored the optical properties of our experimentally synthesised MoS₂ nanostructures with distinct crystal shapes and the number of layers using micro-Raman and photoluminescence (PL) spectroscopy. Figure 2 shows representative Raman spectra of the quasi-1D nanoribbons with widths ranging from 260 to 800 nm, as well as single- and multilayer triangles. The width of the multilayer nanoribbon (annotated with the symbol w in Figure 2a) excludes the contribution from the monolayer edge. The Raman spectrum of the 2D triangle MoS₂ shows two peaks at 384.2 cm⁻¹ and 404.2 cm⁻¹, attributed to the in-plane $E^1_{2g}$ and out-of-plane $A_{1g}$ lattice vibrations of MoS₂, respectively. The estimated peak difference of 20 cm⁻¹ is in good agreement with previous reports on exfoliated or CVD monolayers[40].[?] Remarkably, the quasi-1D nanoribbons exhibit strikingly distinct Raman features compared to the 2D and 3D structures. The $A_{1g}$ peak blue shifts by more than 4 cm⁻¹ compared to the 2D triangle due to interlayer coupling, while the $E_{2g}$ peak redshifts by more than 2 cm⁻¹ from the dielectric screening. Moreover, a pronounced splitting of the $E^1_{2g}$ and $A_{1g}$ Raman modes is observed, particularly for the narrow 1D-nanoribbons. The splitting vanishes for structures with widths of 800 nm and above. These Raman features have never been observed experimentally, and in sharp contrast to the scenario of strained monolayer MoS₂, lattice deformation (strain) and charge transfer (doping) are unlikely to be the sole factors in multilayer nanoribbons. We hypothesise that the pronounced peak splitting comes from a dual contribution from the multilayer nanoribbon core and highly-aligned monolayer nanoribbon edge. The deconvolution of the Raman spectra enables us to identify low-frequency $E^2_g$ mode at 380.6 cm⁻¹ and 384.5 cm⁻¹ and higher frequency modes at 403 cm⁻¹ and 405.4 cm⁻¹ for the single layer and 260 nm-wide multilayer nanoribbon, respectively. This corresponds to a peak difference of 18.5 and 24.8 cm⁻¹, respectively. The peak fitting results are in Supplementary Information Note 6. Regardless of the aforementioned contributions to the Raman spectra, the $E^1_{2g}$ and $A_{1g}$ Raman modes of the nanoribbon core blueshift with decreasing width, while the intensity ratio $A_{1g}/E^1_{2g}$ is significantly reduced compared to the 2D counterpart, likely owing due to the presence of strain in the 1D-nanoribbons. The Raman intensity maps provide a visual distribution of the nanostructures on sapphire. The overlaid rectangles in Figure 2c highlight the low-intensity region associated with single-layer $MoS_2$ and the brighter regions of the quasi-1D and 3D structures. The edges of the nanostructures have a lower intensity than their inner structures, which is consistent with single-layer nanoribbons observed in the SEM images. The Raman spectroscopic features of (AA(A'...))-oriented crystals are different from those of the conventional AB(A..) counterparts due to a greater reduction in the van der Waals gap, which results in a change in the phonon frequencies. Interestingly, we find that the peak positions of the crystals match well those of multilayer triangular crystals with 3R stacking sequence (AA(A'...)-type stacking reported in previous studies.[37] Hence, the phonon vibrations' stacking dependence can be used to confirm the stacking orientation of the triangular crystals.

The PL spectra of as-grown specimens measured at the same locations as the Raman spectra are shown in Figure 2b. The main PL peak of the 2D triangle is located at around 658 nm (1.884 eV), corresponding to the A direct excitonic transition in $MoS_2$, in good agreement with the PL spectra of the quasi-continuous $MoS_2$ layer (Figure S1). No PL emission can be observed from the 3D triangle-shaped structures, which is consistent with our observations of the multilayer structure. This strongly contrasts with the multilayer nanoribbons, which show PL emission comparable to a single-layer $MoS_2$. We also note that the PL peak blue shifts from 658 nm (1.884 eV) to 655 nm (1.896 eV) correspond to a small energy shift of 27 meV, likely due to native strain variations in the as-grown nanostructures.[41]

**Nanoscale image of the quasi-1D $MoS_2$ nanoribbon.** As discussed, multilayer quasi-1D $MoS_2$ nanoribbons exhibit PL emission. $MoS_2$ crystals with controlled stacking configuration (3R) show PL emission for up to six layers, owing to an enhanced spin-orbit coupling.[37] Thus, an idealised stacking orientation can be estimated from the fingerprint PL and Raman spectra of the $MoS_2$ patch. However, such an estimation is not valid for our nanoribbons. The AFM line profiles of the quasi-1D structures reveal a thickness of 16 nm for the 800 um-wide ribbons and 33 nm for the 260 nm-wide ribbons (see Figure S5). Narrow nanoribbons tend to be thicker than broad nanoribbons, indicating various stages of growth. Hence, the enhanced PL emission in our quasi-1D nanoribbons is presumably attributed to the monolayer edges. However, far-field spectroscopy only provides a macroscale picture of the emission of quasi-1D structures. The nm-scale structural information requires nanoscale optical imaging. Here we have performed tip-enhanced photoluminescence (TEPL) with the optical excitation confined to a few nanometers to enable mapping the spatial distribution of exciton emission from the core and edge of the nanoribbons. The correlated AFM and Kelvin probe force microscopy (KPFM) surface scans were used to localise and directly visualise the as-grown 1D nanostructures, as shown in Figure 3a,b. Moreover, the KPFM image intensity is directly related to the work function of the material under study, enabling accurate identification of the number of layers via quantitative image intensity analysis. A careful inspection of the KPFM image reveals that the region of the quasi-1D nanoribbon appears darker than the underlying sapphire substrate, and this can be explained due to a higher work function of $MoS_2$ compared to sapphire. KPFM mapping enables us to resolve the zigzag edges of the monolayer nanoribbon edge while the nanoribbon core is barely visible. This agrees with previous work, which indicates a difference in work function between single- and multilayer of 70 meV.[42] Small dots on the surface of the quasi-1D nanoribbon, clearly distinguishable in the AFM image, have a different colour contrast (hence work function) than the underneath $MoS_2$ layer in the KPFM scan and can probably be due to oxidation of the top layer. Here, KPFM is only used as a qualitative tool, and a direct determination of the work function of the quasi-1D nanoribbons is beyond the scope of the work. The TEPL map acquired using a 532 nm laser reveals a pronounced emission from the nanoribbon edge (blue area in Figure 3d). In contrast, the nanoribbon core appears dark due to an indirect band gap of the thicker $MoS_2$ structure. The overlaid AFM and TEPL images are also shown in Figure 3d. The integrated TEPL spectra from regions marked with rectangular boxes reveal only excitonic emission from the 2D flake (blue-coloured curve) plotted as in Figure 3h. The excitonic emission from the 2D $MoS_2$ flake (located at the left of the image) is slightly red-shifted compared to single-layer nanoribbon, in good agreement with the far-field spectroscopy results, suggesting that local strain can play a role on the observed shift. Maps of the integrated PL peak position and intensity reveal very small spatial variations across the edge of the nanoribbon (Figure S8). Variation in the PL intensity response of the nanoribbon is not specific to the location of edges, and systematic changes have also been observed in single crystals. Here, we attribute these changes to the strain accumulated in the material during the growth because there is a difference between the thermal expansion coefficient of $MoS_2$ and sapphire. An enhanced exciton emission at the nanoribbon edges has also been observed on other nanostructures (see Supplementary Figure S9).

**SHG of quasi-1D $MoS_2$ nanoribbons.** The lack of inversion symmetry in layered TMDs reflects itself in the second-order optical nonlinear response of vdW materials.[43] Hence, second harmonic generation (SHG) can be readily exploited to identify the crystal symmetry in layered TMDs.[44,?] The origins of the strong nonlinear optical response in single-layer TMDs have been extensively discussed in the literature.[36,44–46] Monolayers 2H-$MoS_2$ exhibit an SHG response due to a lack of inversion symmetry. On the other hand, multilayers $MoS_2$ can have distinct symmetry properties. In the AA(A'...)-type stacked $MoS_2$ (2H), the most thermodynamically favourable prototype, the SHG intensity of the Nth-layered $MoS_2$ strongly depends on the layer number. Odd-numbered layers are non-centrosymmetric, retaining a net dipole, whereas even-numbered layers do not. In contrast, 3R-$MoS_2$ belongs to a centrosymmetric space group, and the non-inversion symmetry is preserved in bulk. For the noncentrosymmetric 3R crystal phase of $MoS_2$, a quadratic dependence of the SHG intensity of the number of layers has been predicted due to constructive interference of the nonlinear dipoles.[47]

Figure 3 show the SHG image of $MoS_2$ nanostructures with distinct dimensionalities acquired at a 785 nm pump wavelength. Here, the light is incident on the surface, and the integrated second-harmonic radiation is collected. Small triangles can be identified on the sapphire substrate, likely attributed to singe-layer $MoS_2$, exhibiting a high SHG intensity. The uniform SHG intensity suggests that individual 2D domains are single crystals. Similarly, larger nanostructures with a nearly triangular shape exhibit a uniform SHG intensity (point 2 in Figure 3f). Moreover, a careful examination of the SHG image reveals edge-enhanced SHG for the quasi-1D and occasionally 3D crystals (points 1 and 3 in Figure 3h), which are substantially higher than the central region. A similar trend is observed at the excitation wavelengths of 825, 875 and 1032 nm, with a significantly reduced SHG intensity for the latter (Figure S10). At first glance, the quenching and enhancement of the SHG signal for the empty and full triangles can be attributed to the 2H and 3R stacking orientation, respectively, in good agreement with the AFM images in Figure 1a and previous reports on $MoS_2$ ribbons.? Similarly, edge-enhanced SHG for quasi-1D and aligned 3D structures can be attributed to the corresponding monolayer edges, which lack inversion symmetry. Still, the inner structures do not, as they have a 2H stacking, as discussed previously.[46,48] Surprisingly, the SHG intensity profiles of quasi 1D-$MoS_2$, full and empty triangle-shaped $MoS_2$ crystals in Figure 3g show that the SHG intensities rise abruptly at edges but decrease much slower towards the inner of the nanostructures. In some cases, the intensity drop does not approach the baseline, particularly for the quasi-1D nanoribbons (plot 1 in Figure 3g). Edge-enhanced SHG was previously attributed to distinct edge states due to the translational symmetry breaking in monolayer $MoS_2$.[49] However, the photon energy of 1.57 eV (785 nm) is far from the resonance wavelength of 0.95 eV (1300 nm), at which a pronounced edge-enhancement SHG was observed. Hence, our findings differ significantly from those described earlier.[49,50] Here, an additional factor that can contribute to the edge-enhanced SHG is the broken crystal symmetry due to band bending, in analogy with band bending-induced SHG in other non-centrosymmetric GaAs.[51] Indeed, one can recognize that the edges of the nanostructures form a sharp 1L-ML $MoS_2$ homojunction enabled by sharp modulation in the $MoS_2$ thickness from one layer to a multilayer within a length of several nm. This leads to an interfacial band bending and a strong built-in electric field. Previous work demonstrated the formation of a type-I band alignment with a conduction band offset of 70 meV,[42] attributed to a difference in the electron affinity difference between the mono- and multilayer $MoS_2$.[42] The band offset plays a decisive role in charge transport at the interfaces.[52]

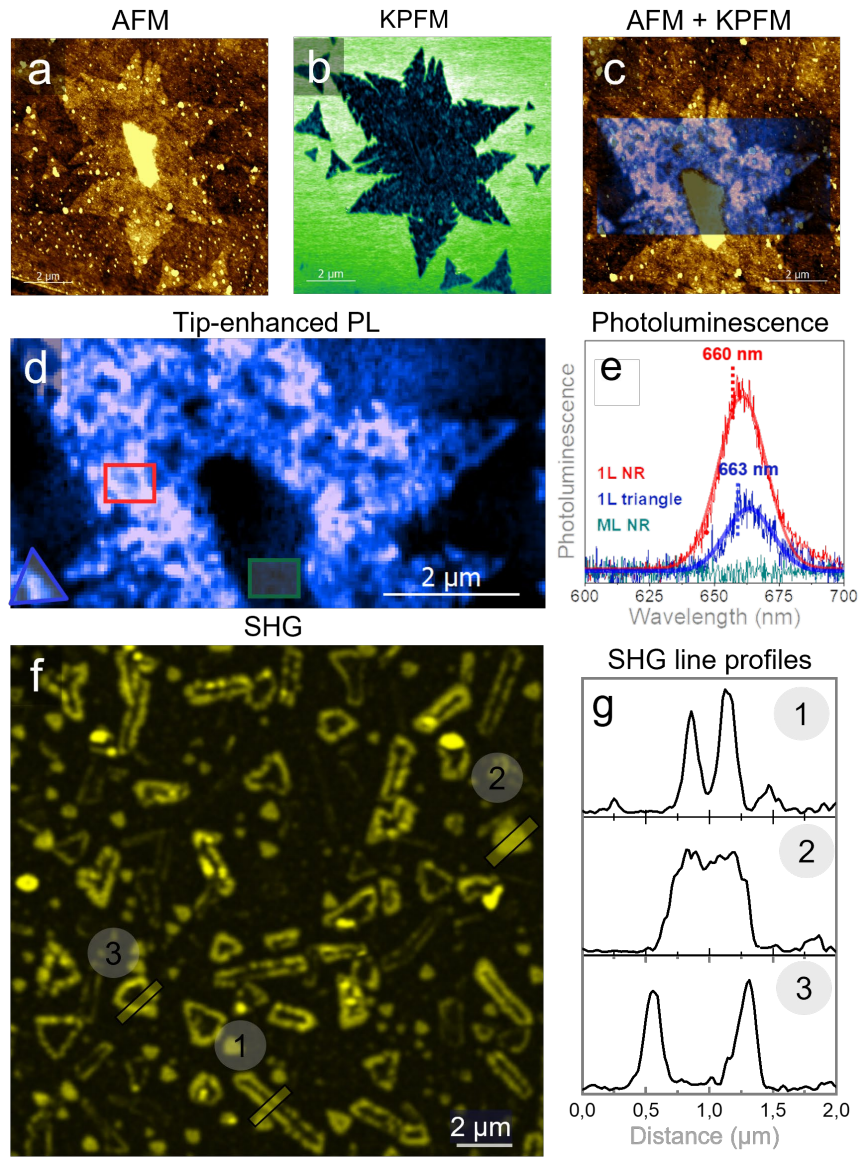

Figure 3: **Nano- and micro-scale imaging of the MoS₂ nanostructures.** (a) AFM topographic image of a MoS$_2$ nanoribbon. (b) Contact potential difference from frequency-modulated Kelvin Probe Microscopy (KPFM). (c) Overlay of the AFM and Tip-enhanced Photoluminescence (TEPL) images. (d) TEPL map of the quasi-1D MoS$_2$ nanoribbon acquired with 532 nm excitation laser. (e) PL spectra averaged from the rectangles overlaid in (d) show the exciton emission from a monolayer nanoribbon (1L nanoribbon), monolayer triangle (indicated as a 1L triangle) and multilayer nanoribbon (indicated as an ML nanoribbon). (f) Second harmonic generation (SHG) image of distinct quasi-1D (1), 2D (2), and multilayer triangles MoS$_2$ (3). (g) Line profiles across the three distinct MoS$_2$ nanostructures, as by the overlaid rectangles in (f).

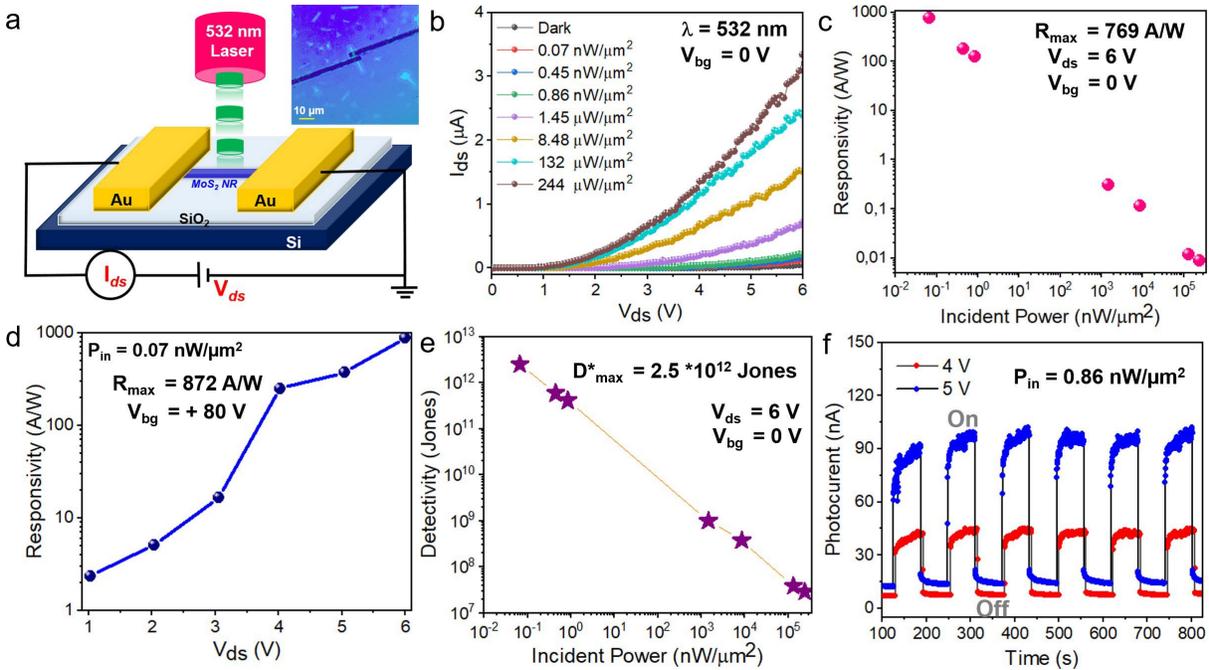

Figure 4: **Single 1D-nanoribbon $MoS_2$ photodetector on $SiO_2$/Si substrate**. (a) Schematic of the quasi-1D $MoS_2$ nanoribbon device with 532 nm laser beam shining on the device channel. The inset shows an optical image of the fabricated device. (b) Output characteristic $I_{ds}$-$V_{ds}$ of the device under dark and various light intensities of 532 nm wavelength. (c) The extracted responsivity of a single quasi-1D $MoS_2$ nanoribbon. The device exhibits a maximum responsivity of 769 A/W at an illumination intensity of 0.07 nW/μm² and $V_{ds}$ = 6 V and $V_{bg}$ = 0 V. (d) Gating responsivity was calculated by applying +80 V on the back gate; the responsivity enhances and reaches the high value of 872 A/W. (e) The detectivity of the single quasi-1D $MoS_2$ nanoribbon was measured at $V_{ds}$ = 6 V and $V_{bg}$ = 0 V. (f) Time-resolved photoswitching response of the device under various $V_{ds}$ of 4 and 5 V.

**Quasi-1D $MoS_2$ photodetectors.** To assess the electrical performance of quasi-1D $MoS_2$ nanoribbons, we fabricated a photodetector device based on a single quasi-1D $MoS_2$ nanoribbon. The photogenerated current under 532 nm illumination was measured by recording the source-drain current ($I_{ds}$) as a function of the applied voltage ($V_{ds}$). Figure 4a illustrates the schematic view of a MoS$_2$ nanoribbon photodetector on a SiO$_2$ (300 nm)/Si substrate. The inset of Figure 4a shows an optical image of the single MoS$_2$ nanoribbon device. Here, Cr/Au lines were used as source and drain contacts, with a spatial separation of 3 µm. As detailed in the device fabrication section, the metal contacts were deposited on top of the MoS$_2$ nanoribbon by standard photolithography. The $I_{ds}$-$V_{ds}$ characteristics of the quasi-1D MoS$_2$ nanoribbon under dark and illumination at 532 nm are shown in Figure 4b. One can observe that there is an almost $10^3$ times increase in photocurrent with increasing power density from 0.07 nW/µm$^2$ to 244 µW/µm$^2$, indicating that the quasi-1D photodetector is highly sensitive to visible light. Indeed, at 244 µW/µm$^2$, the photocurrent in the MoS$_2$ nanoribbon channel reaches nearly 6 $\mu$A ($V_{ds}$=6V), which is three orders of magnitudes higher compared to the dark current (6 nA) at the same applied voltage. Next, we plot the photocurrent ($I_{ph}$) by deducting the ($I_{ds}$) obtained in the dark ($I_{dark}$) from that under illumination ($I_{light}$) as a function of incident power density. The $I_{ph}$ increases sub-linearly with light intensity, as shown in Figure S11. From the output characteristic curves (Figure 4b and Figure S12), the critical figure of merits, i.e., responsivity and specific detectivity were evaluated for the single nanoribbon MoS$_2$ device.

Responsivity ($R_\lambda$) is the photocurrent produced per incident light on an active illuminated area. It is expressed by $R_\lambda$ = ($I_{ph}$)/(PS), where $I_{ph}$ is the photocurrent, P is the incident light density, and S is the illuminated channel area (3.6 $\mu$m$^2$). Figure 4c shows the photoresponsivity of a single MoS$_2$ nanoribbon as a function of incident power ranging from 0.07 nW/µm$^2$ to 244 $\mu$W/µm$^2$ at a fixed applied voltage and without back-gate voltage ($V_{ds}$=6 V, $V_{bg}$=0 V). At a low illumination power density of 0.07 nW/µm$^2$, the device exhibits a remarkably high $R_\lambda$ of around 769 A/W. This value is $10^4$ times higher than one of multi-nanoribbon MoS$_2$ photodetector (45 mA/W),[18] several orders larger than graphene and GaSe nanoribbon devices.[21] The responsivity decreases with increasing effective illumination intensity due to saturation of trap states present at either in MoS$_2$ or at the MoS$_2$/SiO$_2$ interface.[53] Moreover, we observe a small increase in the drain current with the applied gate voltage, that is, from 124 nA ($V_{bg}$=0 V) to 150 nA ($V_{bg}$=80 V) (Figure S12). One can note that the $V_{bg}$ dependence on the photocurrent magnitude is significantly weaker than previous literature reports on MoS$_2$ devices.[18,54] The responsivity of the MoS$_2$ nanoribbon channel at $V_{bg}$=80 V rises substantially from 769 to 872 A/W (Figure 4d).

The other figure of merit, i.e., the specific detectivity (D*), which indicates the device's capability to identify weak optical signals, is expressed by $D^* = RS^{1/2}/(2eI_{dark})^{1/2}$. The calculated D* of our device is $2.5 \times 10^{10}$ Jones at $V_{ds}$=6 V and $V_{bg}$=0 V as shown in Figure 4e. The temporal photoresponse ($I_{ds}$-t) and switching stability of the single MoS$_2$ nanoribbon in response to light illumination were conducted. The $I_{ds}$-t curve was recorded by illuminating the MoS$_2$-nanoribbon device using a 532 nm laser with a continuous ON-OFF cycle under different ($V_{ds}$= of 4 and 5 V) and various illuminations, as shown in Figure 4f and Figure S11. At first, $V_{ds}$ of 4 V and light intensity of 0.86 nW/µm², the device shows a sharp rise in photocurrent under laser ON condition, followed by a sudden drop and slower relaxation in the laser OFF state. This ON-OFF process is continued for several cycles. The reproducibility of consecutive switching ON-OFF cycles confirms the device's robustness and stability. At higher $V_{ds}$ of 5 V, the same observation was noted with a slight increase in the photocurrent, which further proves the reliability of our MoS$_2$ nanoribbon device (Figure 4f). In addition, the ON-OFF measurement was carried out for different powers at constant ($V_{ds}$) = 1 V, clearly demonstrating the stepwise increase of the photocurrent with increasing illumination (Figure S11).

Under equilibrium, without an external bias ($V_g$=0, $V_{ds}$=0), the distribution of charge defects upon illumination is expected to be uniform throughout the nanoribbon core. For simplicity, we exclude the contribution of the nanoribbon edge. As mentioned above, the saturation of trap states occurs at higher illumination, as reflected in the photoresponsivity drop with increasing light intensity (Figure 4c). Under applied voltage ($V_{ds}$ >0), the photo-generated carriers are efficiently swept out under the electric field towards the elec-trodes. This slowly empties the trap states,[53] leading to a significant increase in photocurrent with increasing $V_{ds}$ and more efficiently at higher illumination (Figure 4c). Secondly, the nanoribbon width comparable to the wavelength of light may lead to increased light-matter interaction, thereby increasing photocurrent.[18] The high photoresponse and detectivity could also be attributed to the high crystallinity of the as-grown nanoribbons. Moreover, one cannot exclude the role of the 1L-ML homojunction on the overall increase in the photocurrent of the quasi-1D nanoribbon device. The alignment of the Fermi level at the junction leads to the diffusion of electrons from the monolayer nanoribbon to the nanoribbon core and the accumulation of electrons at the junction, likely resulting in enhanced conductivity along the interface. The band bending at the nanoribbon edge induces a built-in electric field, which can efficiently dissociate excitons and enhance the photocurrent. One can recognize, however, that, given our device geometry, the built-in electric field is perpendicular to the current of the nanoribbon channel and cannot effectively sweep out photogenerated carriers to the contacts. Hence, a $V_{ds}>0V$ bias is required to produce photocurrent in the circuit. Finally,

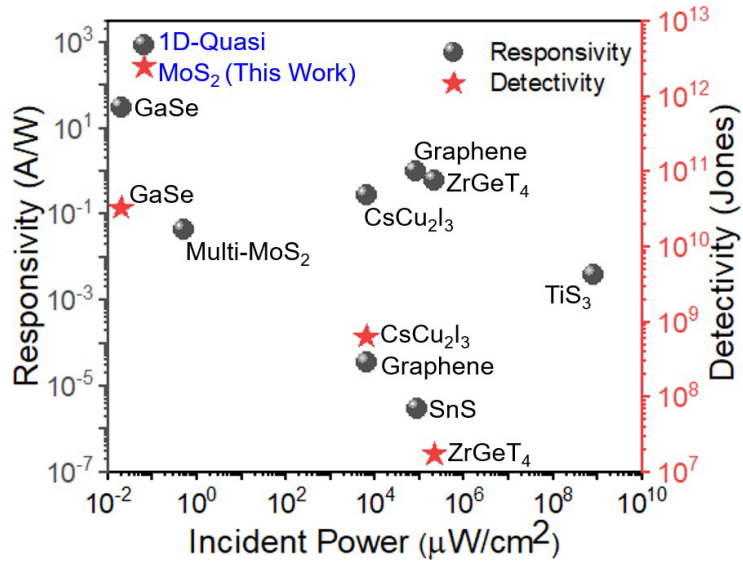

Figure 5: **Overview of various nanoribbon devices' responsivities and detectivities**. Reference data are taken from the literature for SnS,[55] graphene,[20] TiS$_3$,[56] multi-MoS$_2$,[18] ZnSe/Si,[57] CsCu$_2$I$_3$,[58] ZrGeTe$_4$,[3] graphene and reduced oxide graphene,[59] GaSe[21]

we compare the detectivity and responsivity in our quasi-1D MoS$_2$ nanoribbon with that reported for similar devices based on nanoribbons. Figure 5 shows the responsivity (A/W) and detectivity (Jones) as a function of incident power ($\mu$W/cm$^2$). Table 2 of the Supporting Information file also gives an overall comparison. Data for other materials are taken from the literature and correspond to figures of merits of other nanoribbon-based devices, i.e., SnS,[55] graphene,[20] TiS$_3$,[56] multi-MoS$_2$,[18] ZnSe/Si,[57] CsCu$_2$I$_3$,[58] ZrGeTe$_4$,[3] graphene and reduced oxide graphene,[59] GaSe.[21] The attained values are better than any other nanoribbon devices reported up to date and comparable to the commercially available Ge ($10^{11}$ Jones) and Si ($10^{12}$ Jones)-based photodetectors.[60]

## Conclusions

In summary, we have demonstrated the growth of the highly crystalline quasi-1D multilayer (ML) MoS$_2$ nanoribbons with the assistance of NaF as a promotor. The NaF alkali metal halide plays a crucial role in forming multilayer quasi-1D nanoribbons and 3D stacked MoS$_2$ crystals, and in the absence of promoters, quasi-continuous MoS$_2$ layers are obtained. The quasi-1D MoS$_2$ nanoribbons exhibit distinct monolayer (1L) edges forming 1L-ML lateral homojunction due to abrupt variation in the number of layers. Due to edge defects and strain, the 1L edges exhibit Raman and PL features distinct from those of a standard 2D MoS$_2$ triangle. A splitting of the Raman spectra was observed in the quasi-1D MoS$_2$ nanoribbons arising from distinct contributions from the 1L edges and ML core, which become more pronounced with the reduced width of the nanoribbon. Multiphoton microscopy reveals a greatly enhanced optical second harmonic generation (SHG) from the 1L edges of the nanostructures, likely arising due to a non-centrosymmetric 1L and/or symmetry breaking at the surface. Moreover, we report on a high-performance single-nanoribbon MoS$_2$ photodetector a remarkable responsivity of $7.69 \times 10^2$ A/W and measured specific detectivity of $2.5 \times 10^{10}$ Jones at $V_{bg}$=0 V, superior to previously reported nanoribbon based photodetectors. From a broader perspective, this work demonstrates a simple optoelectronic device architecture that can be applied to other TMDs to achieve high responsivity and detectivity and which features a 1L/ML junction intrinsic to the synthesis process and does not involve top-down fabrication techniques, such as laser thinning or layer-by-layer etching that are amenable to the generation of intrinsic defects.

# MATERIALS AND METHODS

## Materials Synthesis

### Growth of MoO$_x$ films by PLD

The precursor oxides were grown by laser ablation of a one-inch MoO$_3$ target (99.95 % purity, from Testbourne Ltd.) in Ar at a pressure of 0.1 mbar. The target was ablated using a 248 nm KrF excimer laser operating at 1 Hz. The target to substrate distance was 7 cm. Before deposition, the vacuum chamber was pumped down to $6 \times 10^{-7}$ mbar base pressure, and the target was pre-ablated using 60 laser pulses. All experiments maintained the laser fluence on the target at 2 J/cm$^2$. The films were grown on (0001) Al$_2$O$_3$ substrates at a temperature of 700°C using several laser shots varying from 5 to 20 pulses.

### Deposition of the NaF layer

The 20 nm-thick NaF layer was deposited on top of the oxide precursors by thermal evaporation of NaF powder (99.9% purity, Sigma Aldrich) in a high vacuum at a pressure of $6 \times 10^{-6}$ mbar. The depositions were done in an evaporation chamber from Univex 250 Oerlikon. The thickness of the NaF layers was estimated using an evaporation rate of 0.3 Å/s.

### Synthesis of the MoS$_2$ nanostructures

The high-temperature growth of MoS$_2$ was carried out in a compact furnace, type OTF-1200X-4-NW-UL, from MTI Corporation. Firstly, the precursor films were placed on a ceramic plate and loaded in the middle of the quartz tube of 4 inches outside diameter. An alumina ceramic boat containing sulfur flakes (1.5 g, purity 99.99, from Sigma Aldrich) was placed outside the central heating zone of the furnace and heated independently using an external heater. The distance between the ceramic boat and the samples was 28 cm. Before sulfurization, the quartz tube was evacuated to $1.6 \times 10^{-3}$ mbar and filled with Ar-5%$H_2$ to 650 mbar to remove residual gases. After repeating the flushing cycle five times, the furnace was heated to 800°C at a rate of 25°C per min. The sulfurization process took place at 800°C for 10 min. When the furnace reached 800°C, the external heating zone containing the sulfur boat reached a temperature of 230°C. When the growth was terminated, the system was cooled down at a rate of 20°C/min to 600°C and then allowed to cool naturally to room temperature. A constant Ar-5% $H_2$ gas flow rate of 100 sccm was maintained throughout the process.

**Materials Characterisation**

Raman and PL spectra were collected using a home-built Raman confocal spectroscopy setup using a 532 nm excitation laser. The spectrometer is a Spectra Pro HRS-750 scanning monochromator from Princeton Instruments equipped with three gratings of 300 gr/mm, 1200 gr/mm, and 1800 gr/mm and a cryogenically-cooled, ultra-low noise Pylon CCD camera, type PyLoN:100BR. The wavelength calibration was carried out using an Ar-Ne light source mounted directly to the entrance slit of the spectrometer. The spectrometer resolution is 1 cm$^{-1}$. Raman and PL spectra were collected using 300 gr/mm and 1800 gr/mm, respectively, and the laser power was kept below one mW to prevent samples from overheating. Leica TCS SP8 CARS microscope (Leica, Germany) was used for second harmonic generation (SHG) imaging of $MoS_2$ nanostructures. The system is equipped with a Pico Emerald laser (APE, Germany), where the stokes laser wavelength is fixed at 1032.4 nm, and the pump laser is tunable in the range of 700-990 nm. Both lasers have a pulse duration of 2ps with a repetition rate of 80 MHz. SHG images were carried out versus pump laser wavelength being tuned to 700, 750, 800 and 850, and 900 nm. Laser power was varied in a range from 12 to 25 mW. Stokes laser at wavelength 1032.4 nm was set to 25 mW of laser power at the sample. The specimens were illuminated through an objective lens with a magnification of 100X (Zeiss Objective EC Epiplan-Apochromat 100x/0.95 HD DIC M27), resulting in a diffraction-limited spot size with a full width of half maximum of around 380-520 nm (depending on laser wavelength). The resulting SHG signals are collected by the same objective as epi-SHG images and via condenser lens (NA = 0.55) as forward-SHG images. All data were collected at a raster speed of 400 Hz using the Galvano scanning head of the Leica TCS SP8 CARS microscope. The high-resolution SEM images were obtained on a Zeiss Merlin microscope with an InLens detector. The nanostructures were imaged using low acceleration voltages (1–2 kV) and short working distances (3 mm). The atomic force microscopy (AFM) images were measured on a Dimension Icon AFM (AFM Icon-PT 2 from Bruker) using Al-coated Si probe tips (type Tap150Al-G from BudgetSensors) The measurements were carried out in non-contact mode in ambient air. The $MoS_2$ samples were transferred onto a TEM grid with the aid of polymethylmethacrylate (PMMA) polymer using a portable transfer method. The TEPL measurements were performed using a nanoRaman system (XploRANano, HORIBA Scientific) integrating an atomic force microscope (SmartSPM) and a Raman microscope (XploRA) with a 100× WD objective tilted by 60° with respect to the sample plane for excitation and collection. A 532 nm excitation p-polarized laser was focused onto the cantilever-based silver-coated AFM-TERS tip (OMNI-TERS-SNC-Ag, Applied Nanostructures Inc.). The true nanoPL map is obtained from recording two PL maps in a special mode called "Spec-TopTM" mode witha "dual spec" option: for each pixel (i) one spectrum (sum of the near-field and far-field signals) is acquired with the tip in direct contact with the surface with a typical interaction force of 2-10 nN and (ii) another spectrum is acquired with the tip a few nm away from the sample surface, considered to be the far-field contribution. In between two pixels of the map, the sample moves in alternating contact to preserve the sharpness and plasmonic enhancement of the tip.

**Preparation of the MoS₂ nanostructure samples on TEM grid by wedging transfer method**

For STEM imaging, MoS$_2$ is transferred from the sapphire substrate to the TEM grid by PMMA coating followed by the KOH etching technique. In this process, the PMMA A4 solution (concentration 10 mg/100 ml in anisole) produces a polymer film with a thickness of 100 nm. The MoS$_2$ sample on sapphire is spin-coated with PMMA at 3000 rpm speed, followed by heating on a hot plate at 120$^0$C for a few minutes. The PMMA layer acts as a handle layer to transfer the MoS$_2$ specimen from the sapphire substrate to the TEM grid. The sample region was then marked and scratched using a diamond cutter under the microscope. The marked area of the sample is usually smaller than the TEM grid. The specimen was then dipped into a 1% potassium hydroxide (KOH) solution for a few minutes. KOH intercalates between the sapphire and sample and easily detaches the PMMA-coated MoS$_2$ from the sapphire substrate. Afterward, to remove KOH residue, the sample was transferred to a Petri dish containing deionized water (DI water), washed several times, and finally picked up by TEM grids. The specimens on the TEM grids were again cleaned in acetone vapor as PMMA can easily dissolve in acetone. After removing the PMMA, the samples were heated at 90ºC to ensure improved adhesion contact between MoS$_2$ and the TEM grid. The presence of the samples on the TEM grid after cleaning with acetone vapor was verified by Raman measurements. Finally, the MoS$_2$ samples were further annealed at 250ºC in an Ar-H$_2$ flow (100 sccm flow rate) for two hours to remove the organic residue. The samples transferred on the TEM grid were baked at a temperature of 160ºC in a vacuum for 8 hours before loading them to the STEM chamber for imaging. The STEM measurements were performed using an aberration-corrected Nion UltraSTEM 100 microscope equipped with a cold field emission gun. The images were acquired at an acceleration voltage of 60 kV and a semi-convergence angle of 31 mrad.

**Device fabrication and Measurements**

Using the above procedure, the as-grown MoS$_2$ nanoribbons were transferred to the Si substrate with a SiO$_2$ thickness of 300 nm at the top surface. The drain-source contacts were patterned via UV photolithography of a Maskless Aligner using a 405 nm laser diode array. The electrical contacts Cr (10 nm)/Au (100 nm) were deposited by electron beam evaporation at a pressure of $1 \times 10^{-6}$ mbar and a depositing rate of 0.5 Å per second. The separation between the source-drain electrodes is 3 $\mu$m. Then the devices were annealed at 300ºC in Ar-5% H$_2$ gas for two hours. The optoelectrical studies were conducted in the ambient environment of room temperature and measurements using a Keithley 2636A semiconductor analyzer under dark and illumination of various intensities of a 532 nm laser source.

**DFT optimizations and STEM simulations**

The MoS$_2$ nanoribbons were modeled by defect-free MoS$_2$ bi- and trilayers with different layer stackings. The different stackings were built based on single 1H-MoS$_2$ monolayers from the Computational 2D Materials Database (C2DB).[61] Different stackings were created by rotation and/or shifting of the layers with respect to each other and optimizing the structure of the multilayer and the cell thereafter. The gap between the multilayers was chosen to be 20 Å. All considered stackings and their theoretical STEM results are given in Fig. S7.

The optimizations were carried out using periodic density functional theory calculations within the projector augmented wave (PAW) method[62] using the atomic simulation environment (ASE)[63] and the GPAW code.[64] The electron density was represented by plane waves with a cutoff of 900 eV. Exchange and correlation were treated by the Perdew-Burke-Ernzerhof (PBE) functional[65] and reciprocal space was sampled with a k-point density of 6 Å$^{-1}$.

The theoretical STEM images and corresponding line profiles of the relaxed atomic structures were simulated using the abTEM code.[66] The parameters of the simulations were set to match the experimental setup, i. e. an acceleration voltage of 60 kV and a semi-convergence angle of 31 mrad.

## Acknowledgements

S.C. acknowledges support from the Independent Research Fund Denmark, Sapere Aude grant (project number 8049-00095B). STEM imaging was conducted as a part of a user project at the Center for Nanophase Materials Sciences, ORNL, a DOE Office of Science User Facility.

The following files are available free of charge.

# Supporting Information

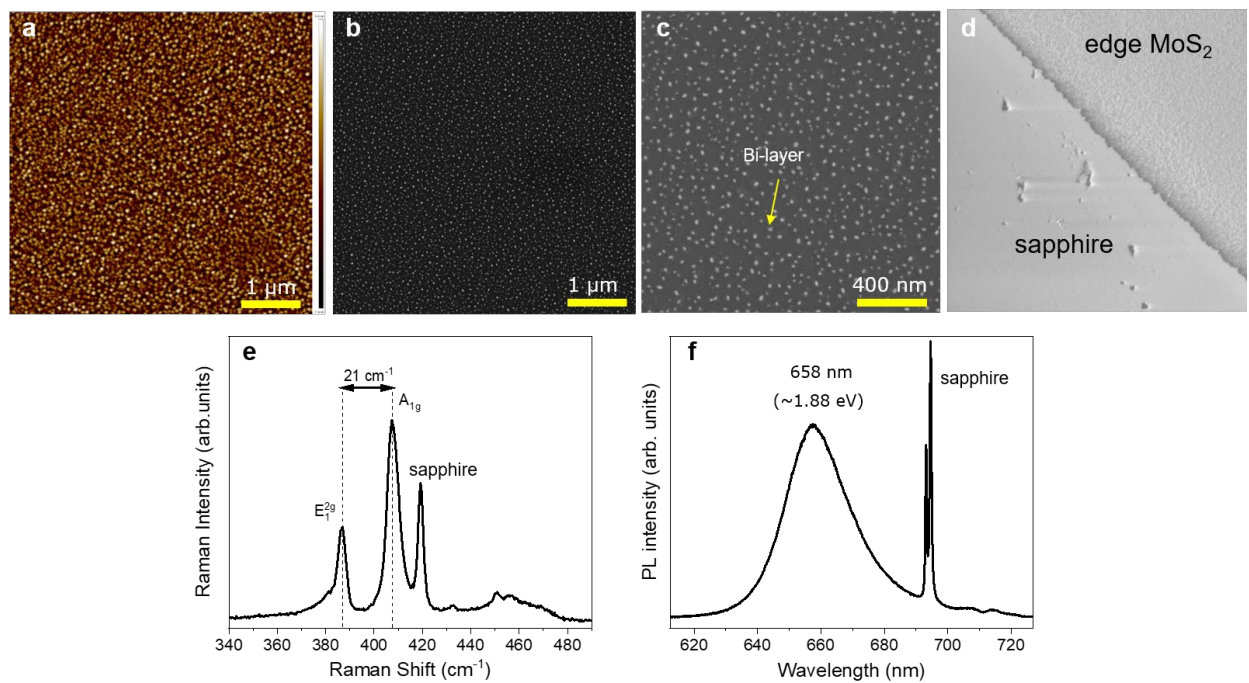

Figure S1: **Quasi-continuous MoS₂ layer synthesized via sulfurization of PLD-grown MoO$_x$ films without NaF**. (a-c) AFM and SEM images showing quasi-continuous MoS₂ film on sapphire. (d) SEM image showing the edge of the MoS₂ layer. (e) Raman spectra of MoS₂ revealing a thickness of 1-2 L. (f) PL spectrum of MoS₂ reveals a crystal quality comparable to a standard CVD process.

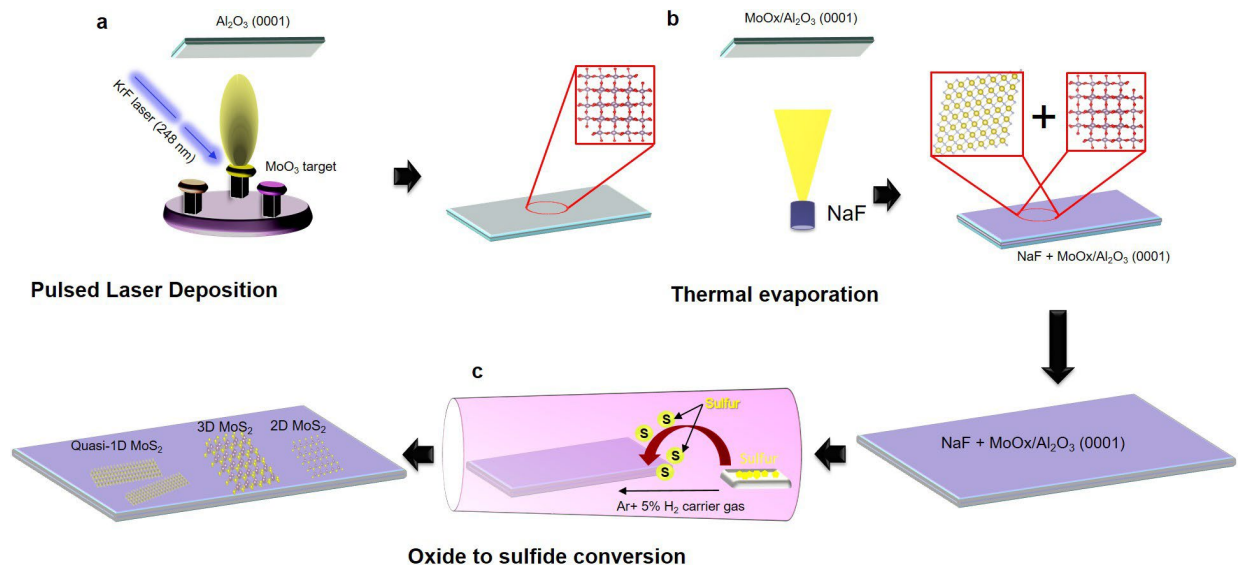

Figure S2: **Pulsed Laser Deposition (PLD)-assisted synthesis of quasi-1D nanoribbons and triangular-shaped MoS$_2$.** a) PLD of epitaxial uniform films MoO$_x$ on sapphire using 20 laser pulses; b) Thermal evaporation of 20 nm NaF at room temperature; c) Oxide to sulfide conversion in a sulfur-rich atmosphere and Ar-H$_2$ gas flow).

Table T1: Peak positions, peak intensities, and intensity ratios of different MoS$_2$ nanostructures.

| MoS$_2$ nanostructure | $E^1_1$ | $E^1_2$ | $A_{1g}$ | $A'_1$ | $\Delta k1$ | $\Delta k2$ | $A_{1g}/E^1_1$ | $A'_{1g}/E^1_{2g}$ |
|---|---|---|---|---|---|---|---|---|
| Quasi-1D nanoribbon (w=260 nm) | 384.5 | 380.8 | 402.9 | 406.5 | 18.4 | 25.7 | 1.92 | 2.79 |
| Quasi-1D nanoribbon (w=360 nm) | 384.1 | 379.27 | 403.0 | 405.4 | 18.9 | 26.1 | 2.49 | 2.64 |
| Quasi-1D nanoribbon (W=800 nm) | - | 380.7 | 406.6 | - | 25.8 | - | 2.51 | - |
| 3D triangle | - | 380.8 | 406.7 | - | 25.9 | - | 2.63 | - |
| 2D triangle | - | 383.5 | 402.9 | - | 19.3 | - | 1.49 | - |

$^a \Delta k1 = A_{1g} - E^1_{1g}$, $\Delta k2 = A'_{1g} - E^1_{2g}$

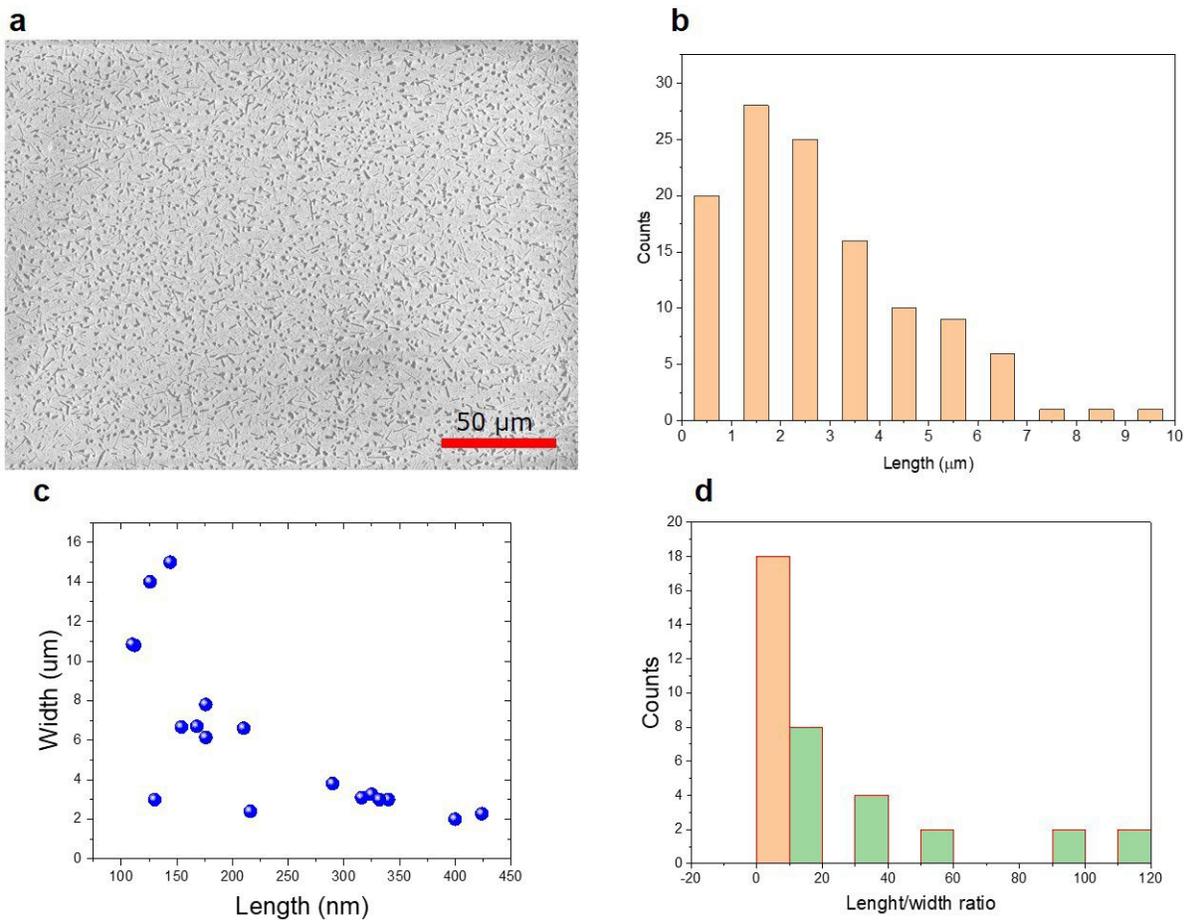

Figure S3: **Size distribution of the quasi-1D MoS$_2$ nanoribbons**. (a) Low magnification SEM image of MoS$_2$ nanoribbons show a high density of isolated domains. (b) Size distribution of the nanostructures reveals a mean length of the nanoribbons in the range of 2 to 3 $\mu$m. (c) Physical dimensions of the nanoribbons, i.e., width versus length. (d) Histogram of the length/width ratios of the nanoribbons.

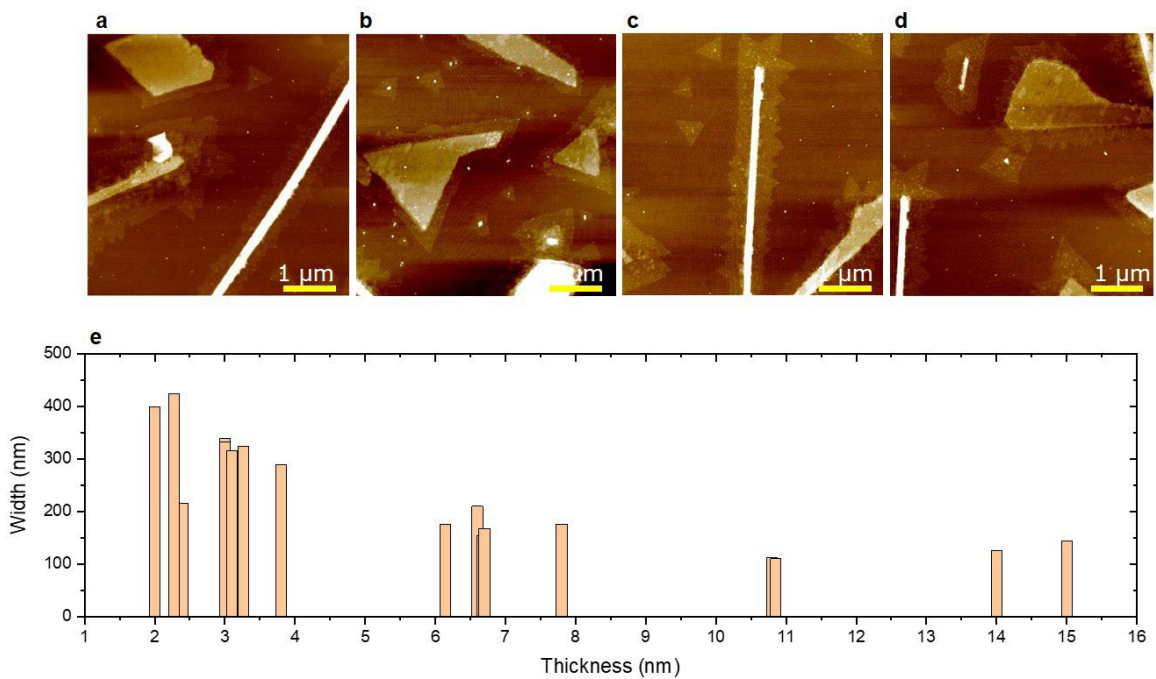

Figure S4: **AFM images and size distribution of various MoS$_2$ nanostructures**. (a-d) Selected AFM images showing quasi-1D nanoribbons and epitaxial 3R-oriented MoS$_2$ triangles. (e) The physical dimensions of the nanoribbons (thickness versus width) are extracted from the AFM images. Narrow nanoribbons are thicker than broad nanoribbons.

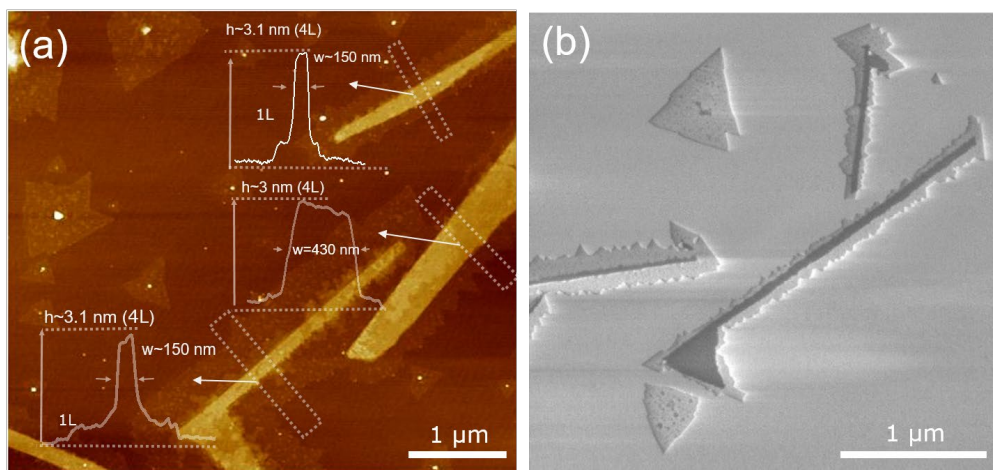

Figure S5: **AFM images of monolayer nanoribbons MoS$_2$ surrounding the multi-layer core**. (a) AFM images showing monolayer nanoribbons surrounding the core structures. Insets show the AFM line profiles extracted from regions highlighted by the dashed rectangles. (b) SEM image showing similar features as in (b) showing monolayer nanoribbons with zigzag edges.

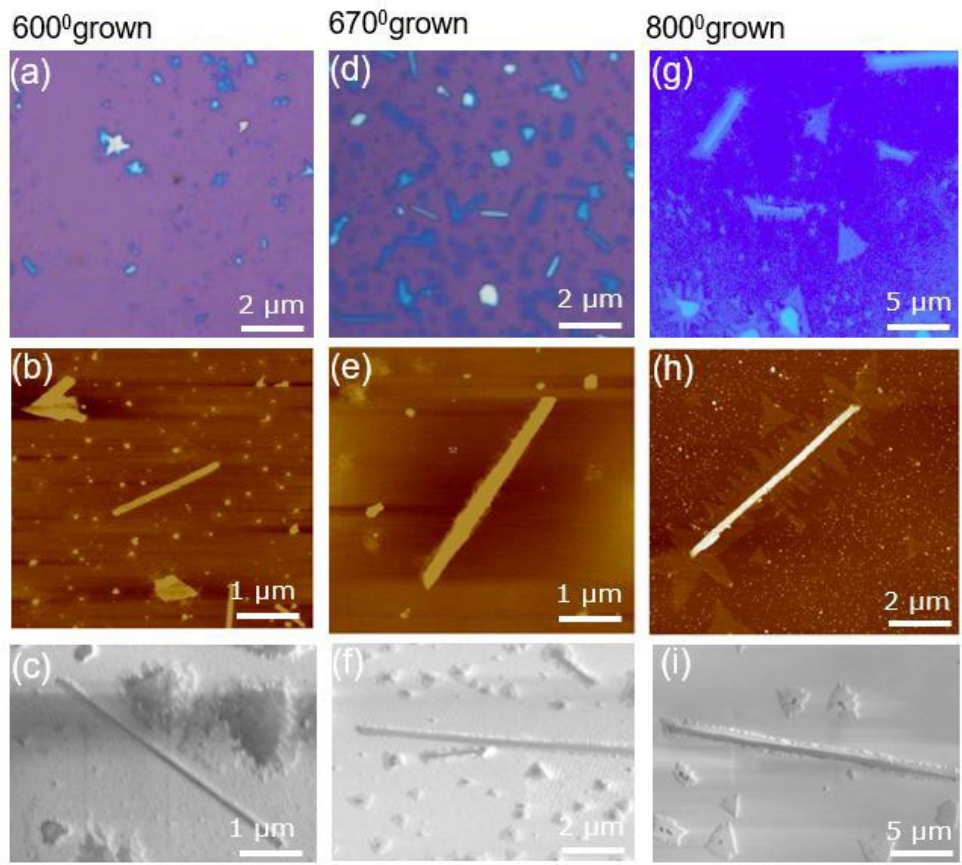

Figure S6: **Temperature-dependence growth of the MoS$_2$ nanoribbons**. (a,d,g) Optical, AFM (b,e,h), and SEM images (c,f, i) of the specimens sulfurized at 600$^0$C, 670$^0$C, and 800$^0$C.

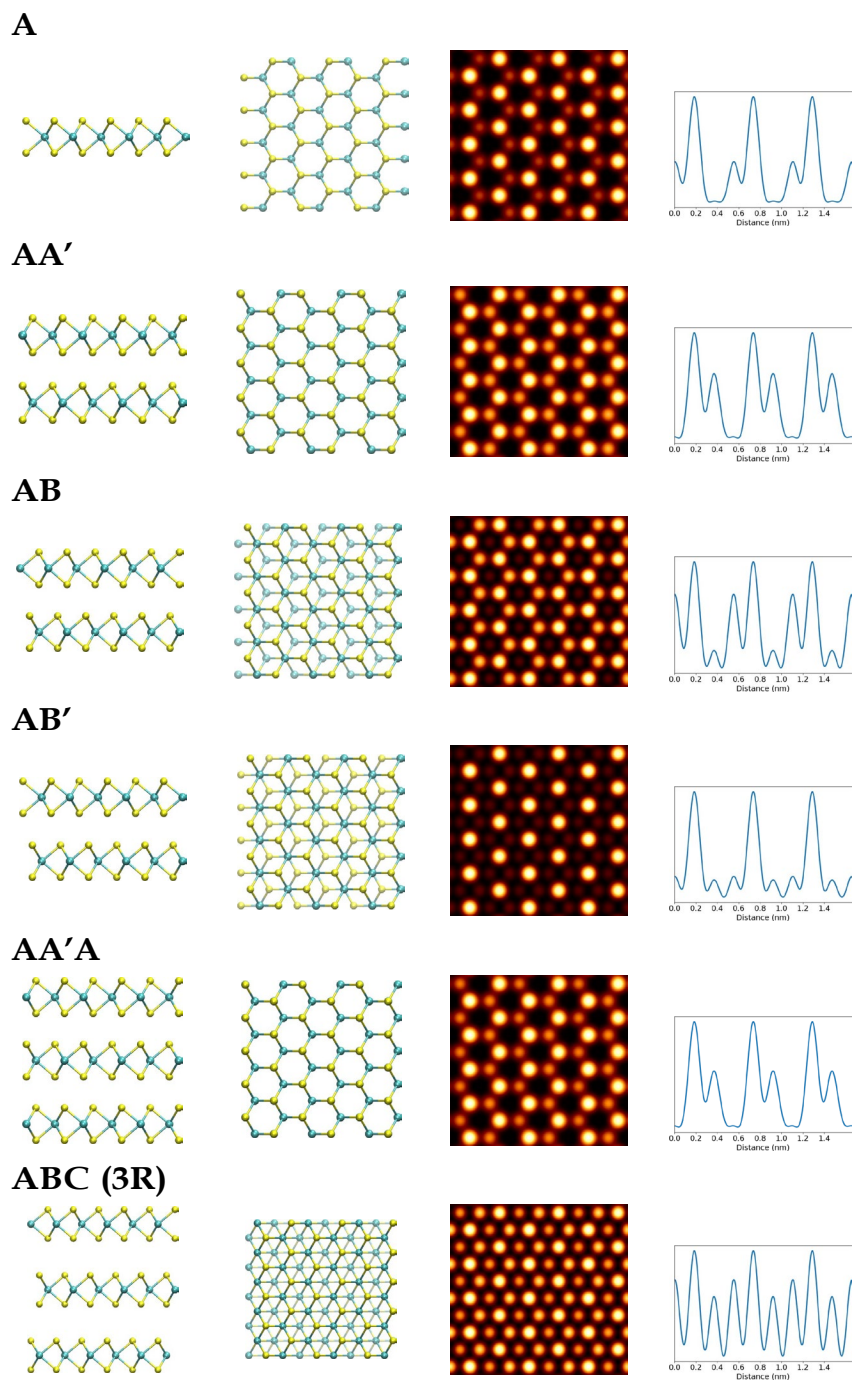

Figure S7: Side view and top view of the optimized MoS$_2$ mono- and multilayers of the considered stackings for modelling the MoS$_2$ nanoribbons as well as their simulated STEM images and line profiles.

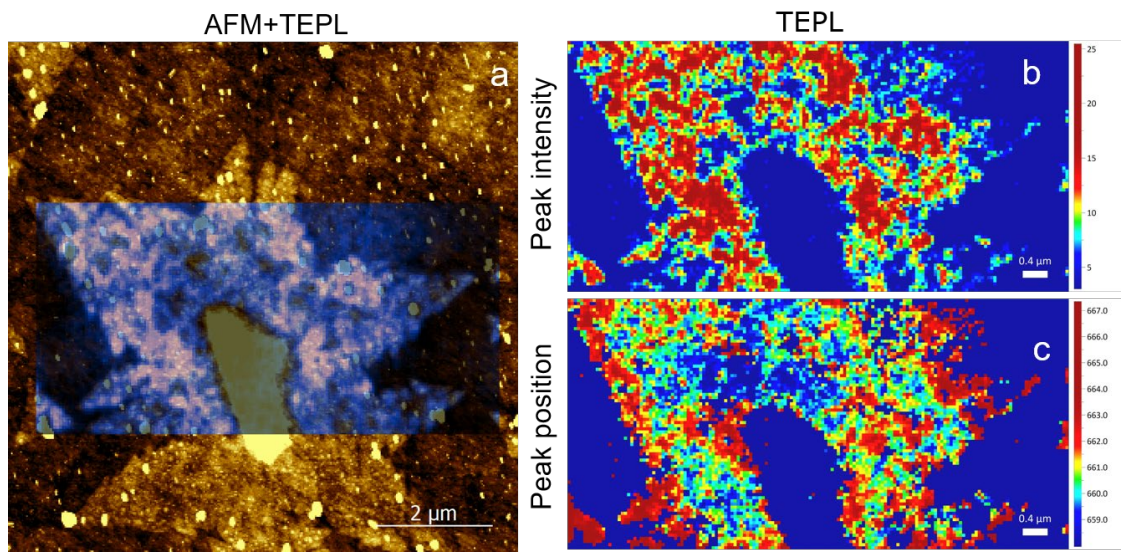

Figure S8: **Tip-enhanced photoluminescence (TEPL) map of the quasi-1D MoS₂ nanoribbon**.(a) Overlap of the AFM and TEPL images of a quasi-1D nanoribbon. TEPL map showing spatial variation of the exciton peak intensity (b) and position across the nanostructure.

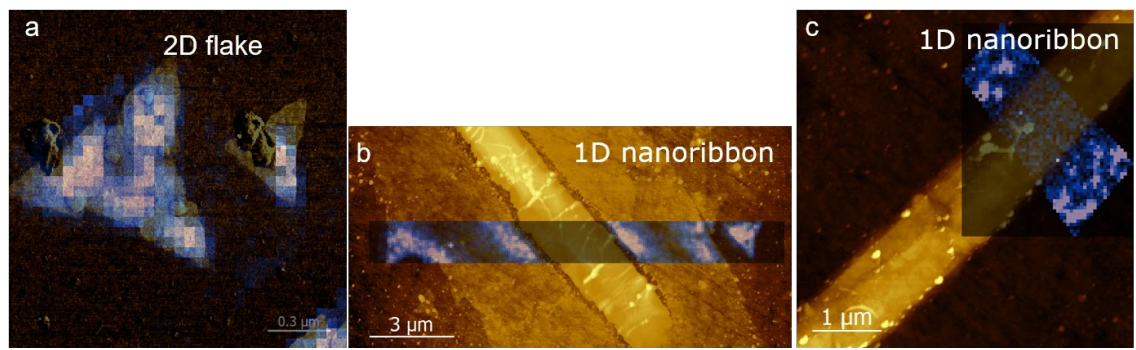

Figure S9: **Tip-enhanced photoluminescence (TEPL) maps**.(a) Overlap of the AFM and TEPL scans of a 2D flake and quasi-1D nanoribbon (b,c).

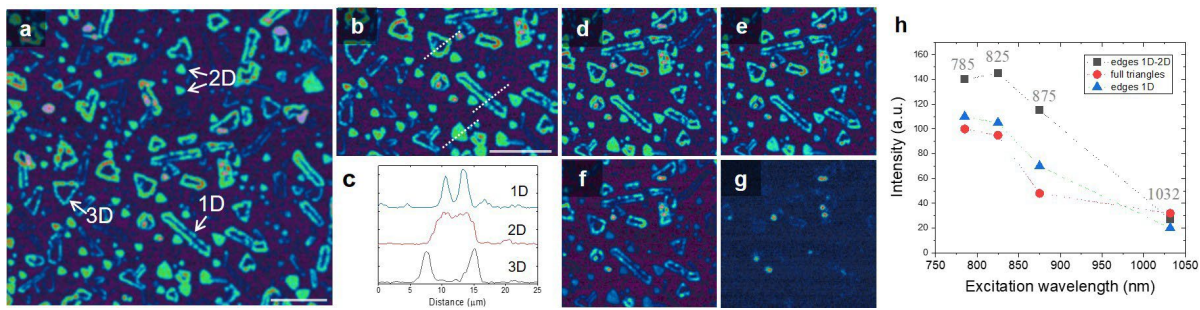

Figure S10: **SHG maps of the quasi-1D MoS$_2$ nanoribbons at 785, 825, 875, and 1032 nm**. (a) SHG map at 785 nm and (b) zoom over a selected area. (c) SHG lines profiles along the lines highlighted in b. (d-g) SHG maps at 785, 825, 875, and 1032 nm. (h) SHG intensity as a function of excitation wavelength for various nanostructures. The data were acquired under similar conditions (laser energy and magnification).

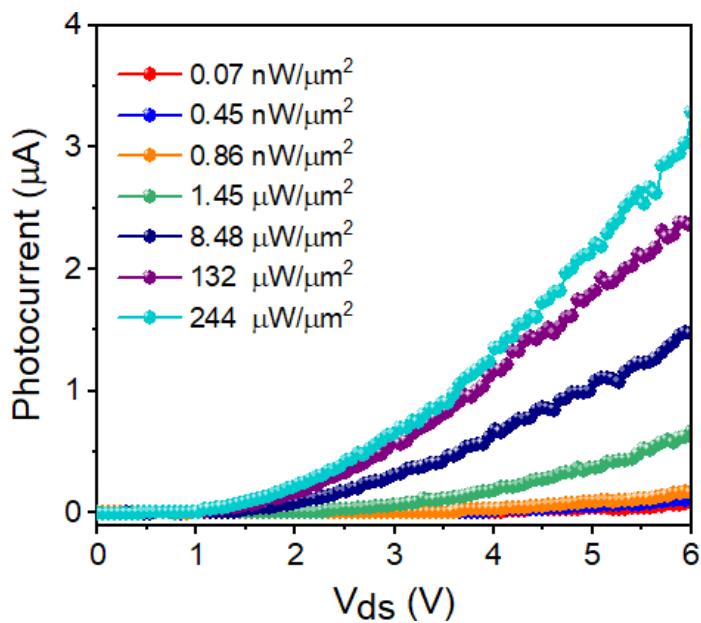

Figure S11: **Photocurrent as a function of applied bias for various illuminations**. Photocurrent in the quasi-1D nanoribbon as a function of bias as a function of illumination.

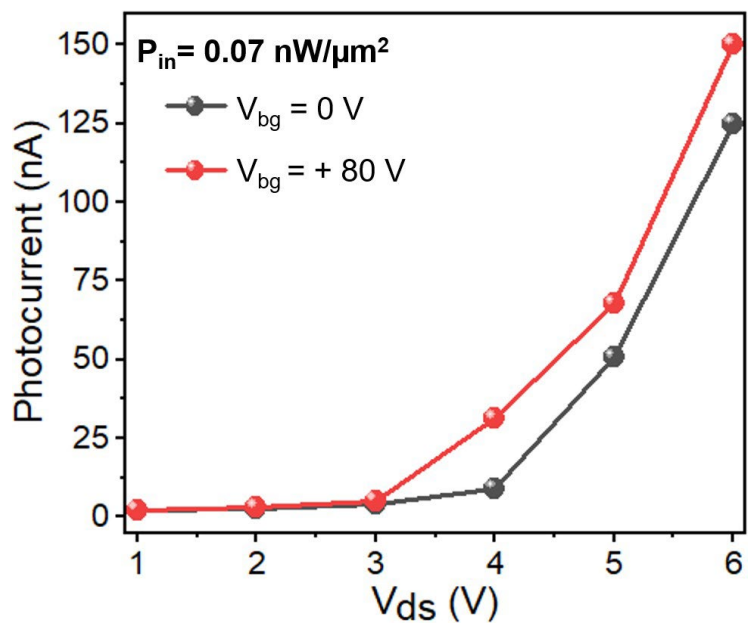

Figure S12: **Gate-dependent photocurrent**.Photocurrent as a function on $V_{ds}$ at $V_g$= 0 V and + 80 V